\documentclass[twocolumn]{aastex631}
\usepackage{amsmath}

\shorttitle{Kepler-33 Transit Timing Variations}
\shortauthors{Sikora et al.}

\graphicspath{{./}{figures/}}

\begin{document}

\title{Refining the Masses and Radii of the Star Kepler-33 and its Five Transiting Planets}

\correspondingauthor{James Sikora}
\email{james.t.sikora@gmail.com}

\author[0000-0002-3522-5846]{James Sikora}
\affiliation{Department of Physics \& Astronomy, Bishop's University, Sherbrooke, QC J1M 1Z7, Canada}
\affiliation{Anton Pannekoek Institute for Astronomy, University of Amsterdam, 1098 XH Amsterdam, The Netherlands}

\author[0000-0002-5904-1865]{Jason Rowe}
\affiliation{Department of Physics \& Astronomy, Bishop's University, Sherbrooke, QC J1M 1Z7, Canada}

\author[0000-0002-6227-7510]{Daniel Jontof-Hutter}
\affiliation{Department of Physics, University of the Pacific, Stockton, CA 95211, USA}

\author[0000-0001-6513-1659]{Jack J. Lissauer}
\affiliation{Space Science \& Astrobiology Division, MS 245-3, NASA Ames Research Center, Moffett Field, CA 94035, USA}

\begin{abstract}
Kepler-33 hosts five validated transiting planets ranging in period from 5 to 41 days. The planets are in nearly co-planar orbits and exhibit remarkably similar (appropriately scaled) transit durations indicative of similar impact parameters. The outer three planets have radii of $3.5\lesssim R_{\rm p}/R_\oplus\lesssim4.7$ and are closely-packed dynamically, and thus transit timing variations can be observed.  Photodynamical analysis of transit timing variations provide $2\sigma$ upper bounds on the eccentricity of the orbiting planets (ranging from $<0.02$ to $<0.2$) and the mean density of the host-star ($0.39_{-0.02}^{+0.01}\,{\rm g/cm^3}$). We combine \emph{Gaia} Early Data Release 3 parallax observations, the previously reported host-star effective temperature and metallicity, and our photodynamical model to refine properties of the host-star and the transiting planets. Our analysis yields well-constrained masses for Kepler-33~e ($6.6_{-1.0}^{+1.1}\,M_\oplus$) and f ($8.2_{-1.2}^{+1.6}\,M_\oplus$) along with $2\sigma$ upper limits for planets c ($<19\,M_\oplus$) and d ($<8.2\,M_\oplus$). We confirm the reported low bulk densities of planet d ($<0.4\,{\rm g/cm^3}$), e ($0.8\pm0.1\,{\rm g/cm^3}$), and f ($0.7\pm0.1\,{\rm g/cm^3}$). Based on comparisons with planetary evolution models, we find that Kepler-33~e and f exhibit relatively high envelope mass fractions of $f_{\rm env}=7.0_{-0.5}^{+0.6}\%$ and $f_{\rm env}=10.3\pm0.6\%$, respectively. Assuming a mass for planet d $\sim4\,M_\oplus$ suggests that it has $f_{\rm env}\gtrsim12\%$.
\end{abstract}

\keywords{{\it Unified Astronomy Thesaurus concepts:} Exoplanets (498); Exoplanet astronomy (486); Transit photometry (1709); Exoplanet dynamics (490); Exoplanet systems (484)}

\section{Introduction}\label{sect:intro}
The Kepler-33 system consists of five short-period ($P<41.1\,{\rm d}$) transiting planets with previously reported radii \citep[derived using \emph{Gaia} Data Release 2 parallax measurements,][]{gaiacollaboration2018} ranging from $1.81\,R_\oplus$ to $4.65\,R_\oplus$ \citep{borucki2011,lissauer2012,berger2018}. The positions of these planets in the period-radius plane are such that they span the radius valley and radius cliff \citep{fulton2017,fulton2018,hsu2019}: planet b is positioned below/close to the valley, planet c ($2.76\,R_\oplus$) falls slightly below the cliff, and planets d, e, and f are positioned above the cliff ($3.47-4.65\,R_\oplus$). Masses previously reported by \citet{hadden2016} based on transit timing variations (TTVs) suggest that planets d, e, f, and likely c all host atmospheres characterized by low mean molecular weights with d exhibiting an anomalously low density of $\approx0.25\,{\rm g/cm^3}$ \citep{chachan2020}. High-resolution spectra of the host-star have been used to derive a solar metallicity, a mass of $\approx1.1-1.3\,M_\odot$, and an age of $\approx4.7\,{\rm Gyrs}$ implying that Kepler-33 will soon be evolving off the main sequence \citep{morton2016,lissauer2012,fulton2018}.

Recently, \citet{hallatt2022} simulated the effects of atmospheric mass-loss for a population of sub-Saturns in order to compare with empirical planet occurrence rates. Their simulations were able to successfully model the majority of known short-period sub-Saturns with available mass constraints. However, based on planet bulk densities, periods, and host-star properties reported in the literature, they identify two planets -- Kepler-223~d and Kepler-33~d -- that likely should have had their atmospheres entirely stripped away. Understanding how these planets could have retained their atmospheres given their host-stars' advanced ages \citep[$\gtrsim4\,{\rm Gyrs}$,][]{morton2016} depends crucially on the accuracy of the adopted planetary and host-star properties.

In this work, we performed a photodynamical modelling analysis (described in Sect.~\ref{sect:transit}) on Kepler-33's light curve in order to derive updated/improved constraints on the radii and masses of the five known planets. This involved re-deriving the host-star's mass and radius using the latest \emph{Gaia} parallax measurements included in the Early Data Release 3 (EDR3) catalog (Sect.~\ref{sect:star}) \citep{gaiacollaboration2021}. Updated mass constraints, presented in Sect. \ref{sect:Mp_Rp}, were obtained for Kepler-33 c, d, e, and f. In Sect.~\ref{sect:composition}, we use the derived masses, radii, orbital periods, and host-star properties to estimate the envelope and core masses of planets e and f. These results, including prospects for further refinement of planet d's mass and atmospheric composition, are discussed and summarized in Sections \ref{sect:disc} and \ref{sect:summary}.

\section{Photodynamical Modelling}\label{sect:transit}

Kepler-33's five transiting planets have orbital periods ranging from $5.66\,{\rm d}$ to $41\,{\rm d}$ \citep{borucki2011,lissauer2012}. Their highly-compact orbits, analogous to the five inner planets known to orbit Kepler-11 \citep{lissauer2011}, are characterized by low inclination angles, low impact parameters, and low eccentricities. The periods of planets c through f place them near various mean motion resonances (MMRs); as a result, planets d, e, and f exhibit relatively large TTVs $\sim10-30\,{\rm min}$, which allowed their masses to be derived by \citet{hadden2016}. Only upper mass limits were reported for planet c while no TTVs associated with planet b were detected.

Photodynamics is the technique of combining photometric modelling of planetary transits with gravitational modelling of interaction between known orbiting planets and the host-star \citep[e.g.,][]{carter2012a}.  The gravitational $N$-body integrator provides time-series 3D positions of the planets and stars which are then used to compute the occurrence of transit events and their photometric properties.  Photodynamics has been successfully used to solve for masses, radii and orbital configurations for dynamically active systems (e.g., \citealt{jontof-hutter2015}) or meaningful upper mass limits (e.g., \citealt{gilbert2020}). 

We used photodynamics to model four years of photometry from NASA's {\it Kepler} Mission.  Photometry products from DR25 \citep{thompson2018} were retrieved from MAST\footnote{https://mast.stsci.edu/portal/Mashup/Clients/Mast/Portal.html}.  Both short-cadence (1-min) and long-cadence (30-min) PDC photometry \citep{stumpe2012,smith2012} were adopted, with a preference for short-cadence photometry when both products are available.  The photometry was normalized by the median flux level for each {\it Kepler} quarter and then stitched together.  The stitched PDF photometric lightcurve had 1,294,087 observations. The photometry was then further processed to filter out stellar and instrumental variability and to remove outliers.  A Savitzky–Golay filter, with a running window of 5-days and a polynomial order of 3 was used for detrending.  The in-transit measurements were excluded from the calculation of polynomial coefficients.  The DR25 best-fit model \citep{thompson2018} was then used to identify photometric outliers using a simple 5-sigma cut with a 5-day running window, resulting in the removal of 846 measurements.  The processed light curve was then visually inspected to verify that identified outliers were not associated with transit observations or other potential astrophysical sources of interest.

The photodynamical model uses the \texttt{TRANSITFIT5} transit modelling software \citep{rowe2015,rowe2016} and the \texttt{Mercury6} hybrid $N$-body integrator \citep{chambers1999}.  The $N$-body model produces positions of each known planet at each observation time-stamp.  The positions are then used by the transit model to calculate the photometric model.  The photodynamical model was parameterized using the mean stellar density $\rho_{\star}$, quadratic limb-darkening, $q_1$, $q_2$ parameterized by \citet{kipping2013}, and a factor to scale the photometric uncertainty reported for {\it Kepler} photometry.  In general, the error reported for {\it Kepler} PDC photometry underestimates the observed point-to-point scatter.  For each planet, the model includes the centre of transit time, T$_0$, defined as when the projected separation between the star and planet as seen by the observer is minimized for the transit closest to the mid-point of the primary {\it Kepler} mission along with the mean orbital period ($P_{\rm mean}$) as observed by {\it Kepler}, the scaled planetary radius ($R_{\rm p}/R_{\star}$), scaled planetary mass ($M_{\rm p}/M_{\star}$), the impact parameter $b_{{\rm T}_0}$, and orbital eccentricity (parameterized by $\sqrt{e}\cos\omega$ and $\sqrt{e}\sin\omega$) observed at T$_0$.

\begin{figure}
	\centering
	\includegraphics[width=1\columnwidth]{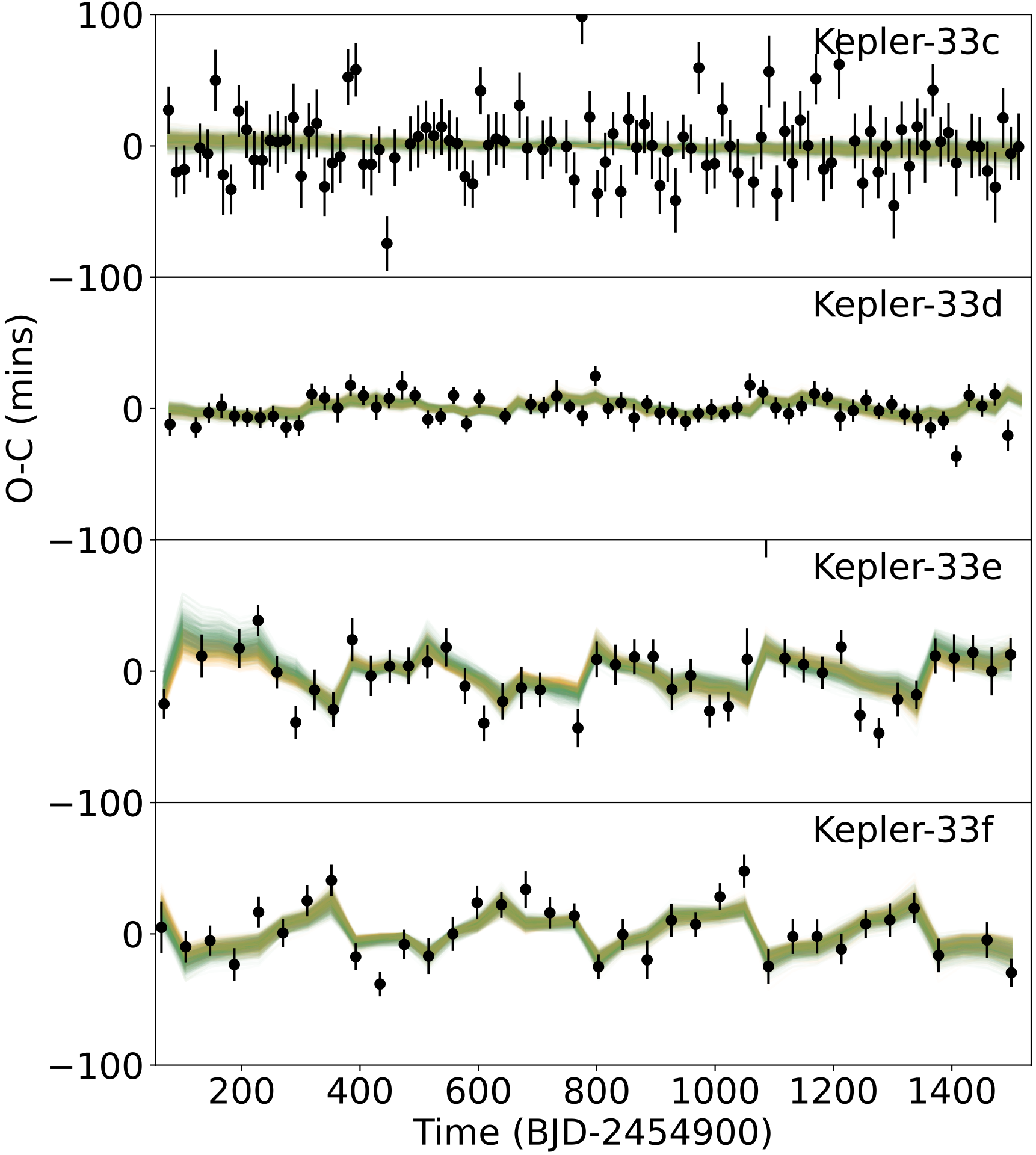}\vspace{0.0cm}
	\caption{O$-$C diagram for planets c, d, e, and f. Black points are associated with the measured transit times as retrieved from DR25 \citep{thompson2018}. Green/yellow lines correspond to the O$-$C values predicted by the photodynamics analysis.}
	\label{fig:ttv}
\end{figure}

\begin{figure*}
	\centering
	\includegraphics[width=1\columnwidth]{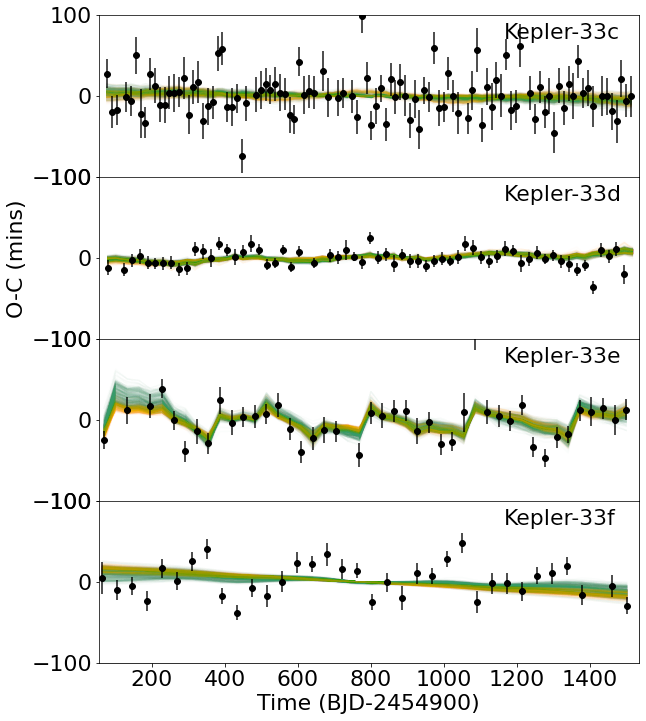}\vspace{-0.0cm}
	\includegraphics[width=1\columnwidth]{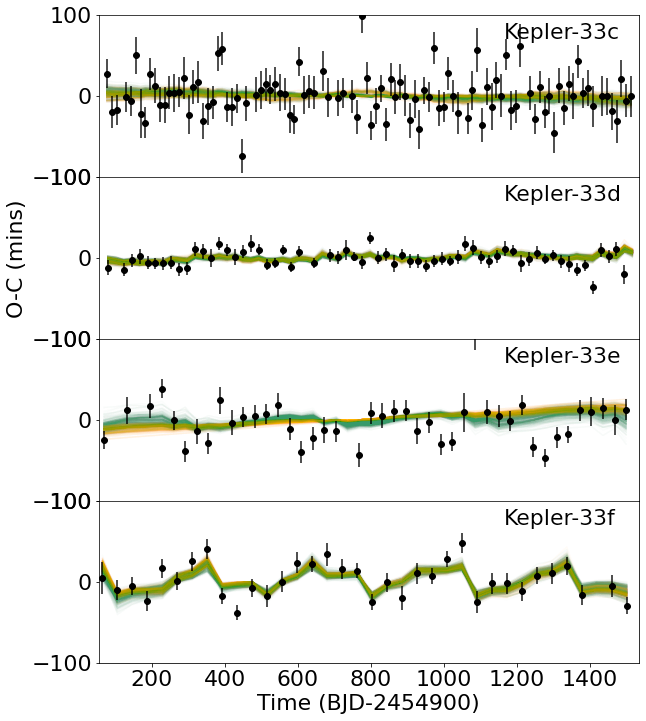}\vspace{-0.0cm}
	\caption{Same as Fig. \ref{fig:ttv} but showing the TTV contributions from planet e (left) and planet f (right) by setting the mass of planet e and f to zero, respectively, and recalculating the models.}
	\label{fig:ttv_pl}
\end{figure*}

\begin{figure}
	\centering
	\includegraphics[width=0.8\columnwidth]{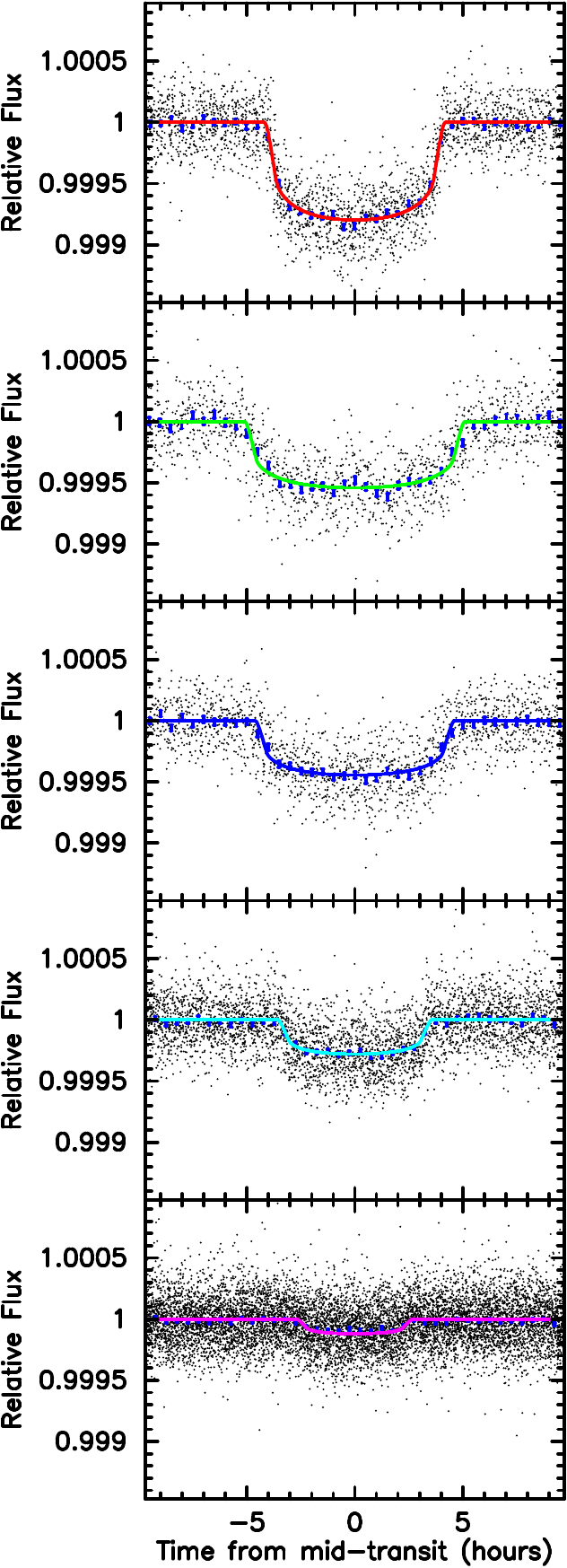}\vspace{-0.1cm}
	\caption{Comparison between the best-fitting transit models (solid lines) and the observed \emph{Kepler} DR25 measurements (black points; blue points are binned values). The panels are ordered from the largest to the smallest planets (from top to bottom: d, f, e, c, b). For each panel, the TTV associated with each transit has been removed along with the contribution from every other planet.}
	\label{fig:phased}
\end{figure}

Model parameters and posterior distributions were calculated using an affine-invariant ensemble sampler with 400 walkers \citep{foreman-mackey2013a}.  Parameters were initialized using the DR25 transit model from Table 7 of \citet{thompson2018} by drawing random models from the published MCMC chains \citep{thompson2018}.  As the DR25 models assume circular, non-interacting orbits a wide range of orbital eccentricity and planetary mass was adopted for $M_{\rm p}/M_{\star}$, $\sqrt{e}\cos\omega$ and $\sqrt{e}\sin\omega$.  The chains were evolved to produce a length of $250,000$.  The first 20,000 steps were discarded as burn-in and the resulting chains were used to calculate posterior distributions as reported in Table \ref{tbl:pl_param}.  A uniform prior for $M_{\rm p}/M_{\star}$ for planet b was used (approximately corresponding to $M_{\rm p}\in(0,50)\,M_\oplus$) in order to remove clearly unrealistic/non-physical high-mass solutions. Convergence was tested using the Gelman-Rubin statistic \citep{gelman1992} applied to each parameter, which yielded $\hat{R}$ values ranging from 1.01 to 1.05. 

Observed minus calculated (O$-$C) transit times are shown for planets c, d, e, and f in Fig. \ref{fig:ttv}. The contributions to the TTVs caused by planets e and f are shown in Fig. \ref{fig:ttv_pl}. The small amplitude of the theoretical TTVs of planets e and f when the mass of the other planet is set equal to zero to zero shows that any TTVs induced by planet d must be close to or less than the detection limit. The best-fitting transit models are plotted in Fig. \ref{fig:phased} and compared with the observed measurements phased by the orbital periods.

\section{Host-Star Properties}\label{sect:star}

High-resolution optical spectra of Kepler-33 were previously obtained as part of the California-\emph{Kepler} Survey (CKS) \citep{petigura2017} using the Keck-HIRES instrument. Based on spectroscopic modelling of these observations, the authors report an effective temperature and iron abundance of $T_{\rm eff}=5947\pm60\,{\rm K}$ and ${\rm [Fe/H]}=0.142\pm0.040$, respectively. \citet{fulton2018} derived a stellar radius of $R_\star^{\rm CKS}=1.609_{-0.045}^{+0.047}\,R_\odot$ using the \emph{Gaia} Data Release 2 parallax \citep{gaiacollaboration2018} of $0.851\pm0.015\,{\rm mas}$ \citep[a correction of $+0.053\,{\rm mas}$ has been applied based on][]{zinn2019} in conjunction with the 2MASS $K_{\rm s}$ magnitude \citep[$12.591\pm0.022$,][]{cohen2003,skrutskie2006}.

We used the publicly available \emph{Gaia} Early Data Release 3 (EDR3) catalog \citep{gaiacollaboration2021} in order to re-calculate $R_\star$. This was carried out using the \texttt{isoclassify} Python package\footnote{https://github.com/danxhuber/isoclassify} \citep{huber2017,berger2020}. The code uses bolometric corrections calculated for the MESA Isochrones and Stellar Tracks (MIST) grid of stellar evolution models \citep{choi2016} and dust maps to account for reddening that are incorporated into the \texttt{mwdust} Python package\footnote{https://github.com/jobovy/mwdust} \citep{bovy2016}. \citet{lindegren2021} provide zero-points to correct for the \emph{Gaia} EDR3 parallax bias\footnote{https://gitlab.com/icc-ub/public/gaiadr3\_zeropoint/-/tree/master}, which, for Kepler-33, is found to be $Z_5=-0.026\,{\rm mas}$. \citet{zinn2021} provide a further refinement of this correction based on \emph{Kepler} asteroseismic measurements of brighter stars, which suggest that an additional correction of $\Delta Z=-0.038\,{\rm mas}$ is warranted along with an increase in the uncertainty of $22\%$. Based on these corrections, we adopt a parallax of ${\hat{\varpi}_{\rm EDR3}}-Z_5+\Delta Z=0.802\pm0.014\,{\rm mas}$, where ${\hat{\varpi}_{\rm EDR3}}=0.814\pm0.012\,{\rm mas}$ corresponds to the raw \emph{Gaia} EDR3 parallax. Reddening of the $K_{\rm s}$ magnitude was estimated using the \citet{green2019} dust map. Using the CKS values of $T_{\rm eff}$ and ${\rm [Fe/H]}$, the bias-corrected \emph{Gaia} EDR3 parallax, and the $K_{\rm s}=12.591\pm0.022$ measurement, \texttt{isoclassify} yields a radius of $R_\star=1.721_{-0.053}^{+0.055}\,R_\odot$. This corresponds to an increase of $7\%$ ($2.4\sigma$) relative to $R_\star$ reported by \citet{fulton2018}, while the uncertainties in both values are comparable.

For this study, additional host-star properties (e.g., mass, age, density) were derived by fitting the MIST grid of stellar evolution models \citep{choi2016} to the CKS $T_{\rm eff}$ and ${\rm [Fe/H]}$ values along with $R_\star$. This was done using the \texttt{emcee} Ensemble Sampler \citep{foreman-mackey2013a} with 50 walkers taking 100,000 steps each with the first 10,000 iterations being discarded as burn-in. The Gelman-Rubin statistic ($\hat{R}$) \citep{gelman1992} was used to test for convergence; all of the resulting chains were considered to have converged based on the calculated $\hat{R}<1.01$. The set of model parameters associated with each sample were computed using the \texttt{isochrones} Python package\footnote{https://isochrones.readthedocs.io/en/latest/}, which allows the MIST model grid to be interpolated based on a given ${\rm [Fe/H]}$, initial mass, and so-called equivalent evolutionary point (EEP). Two priors that are included in the \texttt{isochrones} package were adopted: for the mass, we used the Chabrier broken power-law \citep{chabrier2003}, and for the age, we used Eqn. 17 of \citet{angus2019} (n.b., similar results were obtained using a flat age prior).

\begin{figure}
	\centering
	\includegraphics[width=1\columnwidth]{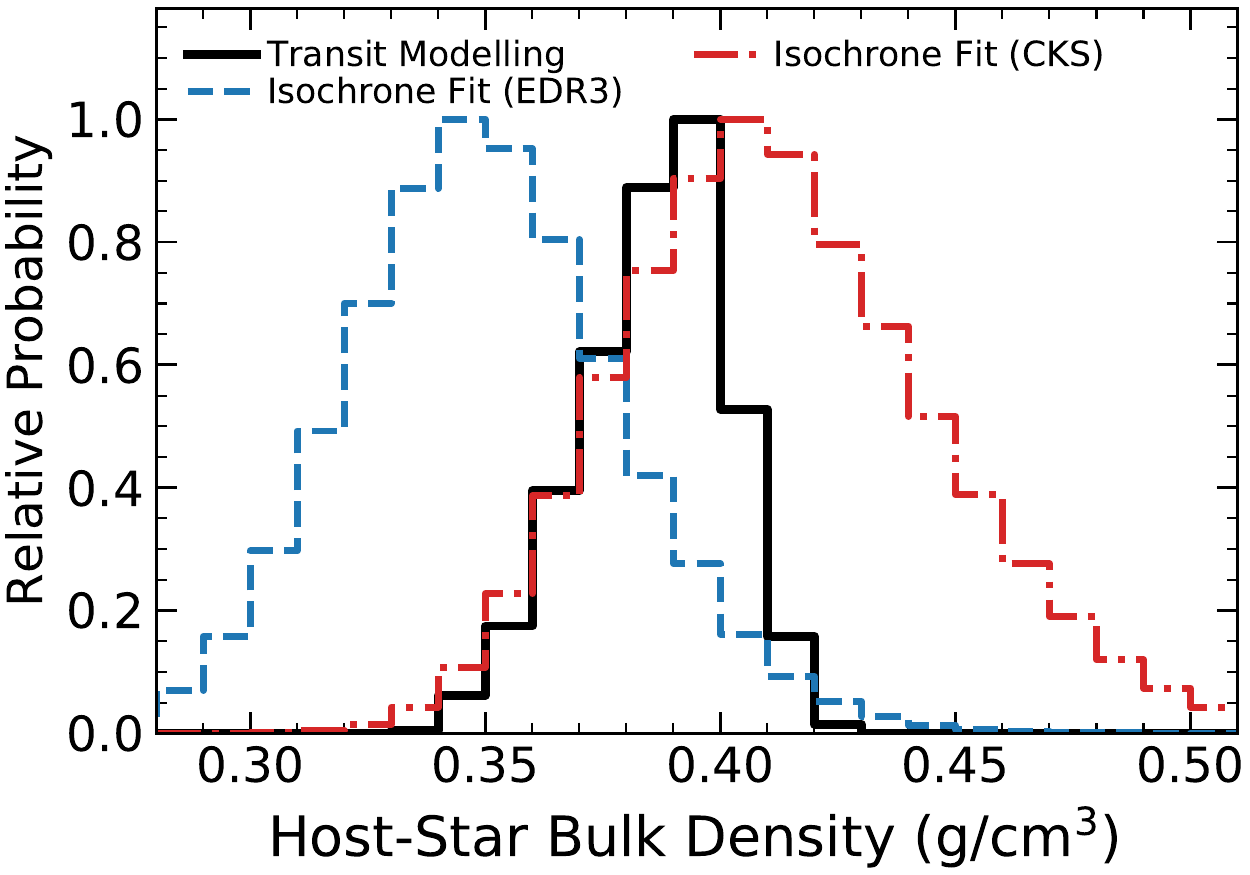}\vspace{-0.1cm}
	\caption{Stellar bulk density posterior distributions computed from the transit modelling (solid black) and from fitting the observed stellar parameters to the MIST grid of evolution models \citep{choi2016}: The dot-dashed red distribution is obtained using the CKS $T_{\rm eff}$, ${\rm [Fe/H]}$, and $R_\star^{\rm CKS}$ values \citep{fulton2018}, while the dashed blue distribution is obtained using the CKS $T_{\rm eff}$ and ${\rm [Fe/H]}$ values along with the newly-derived $R_\star^{\rm EDR3}$ computed using the corrected \emph{Gaia} EDR3 parallax.}
	\label{fig:rho_s}
\end{figure}

Several variations of the MCMC analysis were carried out in which either the new $R_\star$ value derived with ${\hat{\varpi}_{\rm EDR3}}$ ($R_\star^{\rm EDR3}$) or the CKS $R_\star$ value ($R_\star^{\rm CKS}$) were used. We find that in both instances, the derived stellar masses ($M_\star$) differ with respect to the value reported by \citet{fulton2018} ($M_\star^{\rm CKS}=1.208_{-0.065}^{+0.033}\,M_\odot$) by $\sim1.6\sigma$ while the stellar ages ($\log_{10}{\rm Age}^{\rm CKS}/{\rm yrs}=9.68_{-0.06}^{+0.11}$) agree within $\sim0.8\sigma$. In Fig.~\ref{fig:rho_s}, we compare the two derived $\rho_\star$ posterior distributions with that derived from the transit modelling ($\rho_\star^{\rm tr}=0.39\pm0.02\,{\rm g/cm^3}$). The $\rho_\star$ values derived using $R_\star^{\rm EDR3}$ ($0.35\pm0.03\,{\rm g/cm^3}$) and $R_\star^{\rm CKS}$ ($0.40_{-0.03}^{+0.04}\,{\rm g/cm^3}$) differ from $\rho_\star^{\rm tr}$ by approximately $2.7\sigma$ and $0.7\sigma$, respectively.

\begin{table}
	\caption{Properties associated with the host-star derived using $T_{\rm eff}$ and ${\rm [Fe/H]}$ \citep[CKS,][]{fulton2018} along with $R_\star^{\rm EDR3}$ and $\rho_\star^{\rm tr}$; the $\rho_\star^{\rm iso}$ value corresponds to the adopted stellar density derived from the isochrone fit (Sect. \ref{sect:star}) and $\rho_\star^{\rm tr}$ is derived from the transit modelling (Sect. \ref{sect:transit}). $u_1$ and $u_2$ are the quadratic limb darkening parameters. Uncertainties correspond to $1\sigma$.}{\vskip-0.3cm}
	\label{tbl:st_param}
	\begin{center}
	\begin{tabular}{@{\extracolsep{\fill}}l r@{\extracolsep{\fill}}}
		\hline
		\hline
		\noalign{\vskip0.5mm}

\multicolumn{2}{c}{Kepler-33} \\
\hline

$K{\rm p}\,({\rm mag})$ & 13.988 \vspace{0.1cm} \\
$T_{\rm eff}\,({\rm K})$ & $5947\pm60^\dagger$ \vspace{0.1cm} \\
${\rm [Fe/H]}$ & $0.14\pm0.04^\dagger$ \vspace{0.1cm} \\
$R_\star\,(R_\odot)$ & $1.66\pm0.03$ \vspace{0.1cm} \\
$M_\star\,(M_\odot)$ & $1.26_{-0.06}^{+0.03}$ \vspace{0.1cm} \\
${\rm Age}\,({\rm Gyrs})$ & $4.2_{-0.3}^{+1.3}$ \vspace{0.1cm} \\
$L_\star\,(L_\odot)$ & $3.1_{-0.1}^{+0.2}$ \vspace{0.1cm} \\
$\rho_\star^{\rm iso}\,({\rm g/cm^3})$ & $0.38_{-0.01}^{+0.02}$ \vspace{0.1cm} \\
\hline
$\rho_\star^{\rm tr}\,({\rm g/cm^3})$ & $0.39_{-0.02}^{+0.01}$ \vspace{0.1cm} \\
$u_1$ & $0.50_{-0.10}^{+0.12}$ \vspace{0.1cm} \\
$u_2$ & $0.0\pm0.2$ \vspace{0.1cm} \\

		\noalign{\vskip0.5mm}
		\hline
$^\dagger$\citet{petigura2017}
	\end{tabular}
	\end{center}
\end{table}

\begin{figure*}
	\centering
	\includegraphics[width=2\columnwidth]{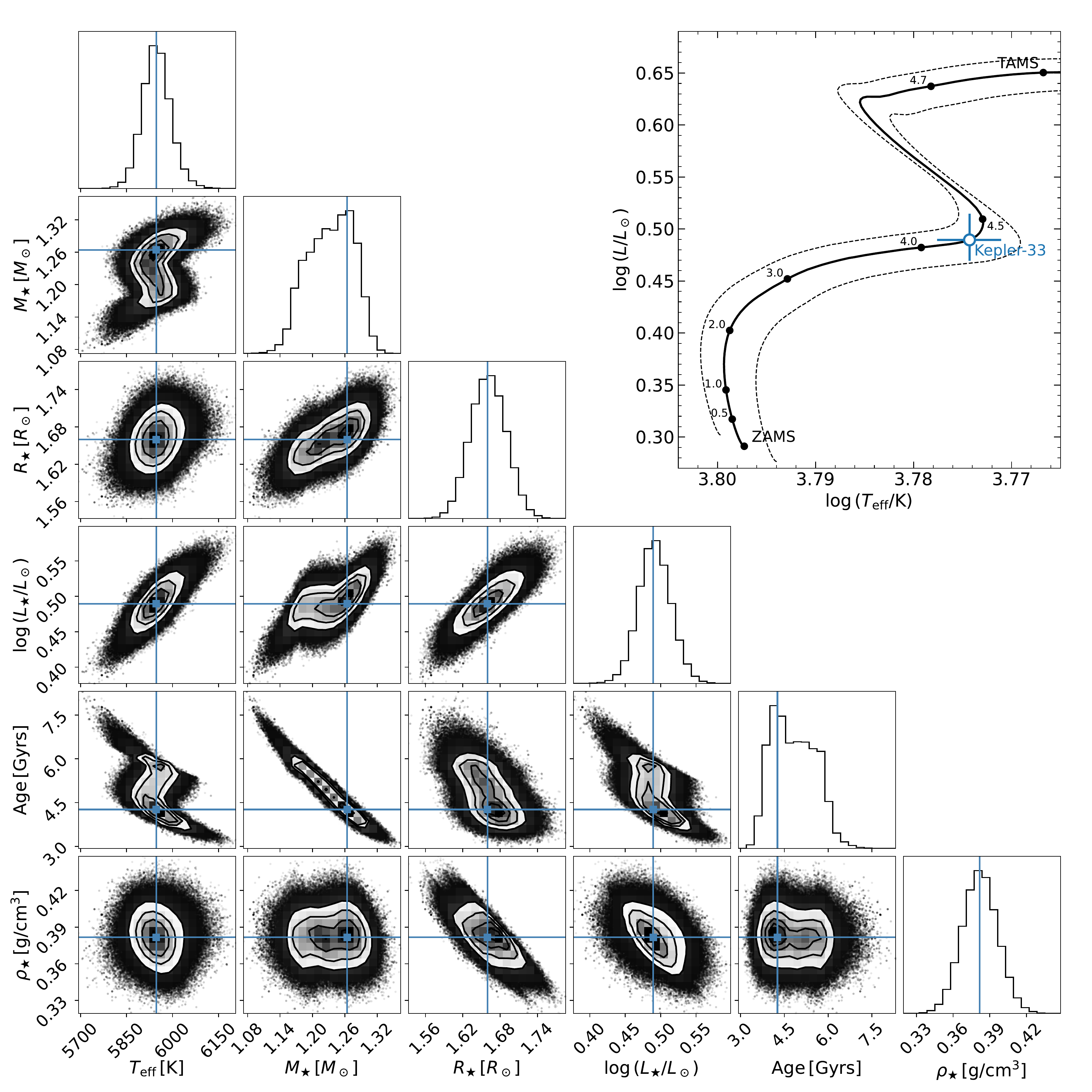}\vspace{-0.4cm}
	\caption{Stellar parameter marginalized posterior distributions generated by fitting Kepler-33's $T_{\rm eff}^{\rm CKS}$ and ${\rm [Fe/H]}^{\rm CKS}$ \citep{fulton2018} along with $R_\star^{\rm EDR3}$ and $\rho_\star^{\rm tr}$ using the MIST grid of evolutionary models \citep{choi2016}. Blue lines indicate each distribution's mode corresponding to the adopted value. The Hertzsprung-Russell Diagram (inset, top right) shows Kepler-33's position (blue open circle) along with the interpolated MIST evolutionary track associated with the adopted parameters (solid black). Black circles indicate the zero-age and terminal-age main sequence (ZAMS and TAMS) along with intermediary points labeled according to the age in Gyrs. Dashed lines correspond to tracks computed for the $1\sigma$ ${\rm [Fe/H]}$ limits of $0.10$ and $0.18$.}
	\label{fig:HRD_corner}
\end{figure*}

\begin{table*}
	\caption{Planetary parameters associated with Kepler-33 b, c, d, e, and f derived in this work; uncertainties correspond to $1\sigma$ while lower/upper limits correspond to $2\sigma$. The parameters in rows labeled from $P_{\rm orb}$ to $M_{\rm p}/M_{\rm s}$ are derived from the photodynamical modelling (Sect. \ref{sect:transit}); $F_{\rm p}$, $R_{\rm p}$, $M_{\rm P}$, and $\rho_{\rm p}$ are derived using the host-star properties shown in Table \ref{tbl:st_param}; the remaining rows list parameters associated with the H/He envelopes for planet e and f, which are derived using the grid of models described by \citet{lopez2014}.}{\vskip-0.6cm}
	\label{tbl:pl_param}
	\begin{center}
	\begin{tabular}{@{\extracolsep{\fill}}l c c c c r@{\extracolsep{\fill}}}
		\hline
		\hline
		\noalign{\vskip0.5mm}
 & b & c & d & e & f \\
\hline

$P_{\rm orb}\,({\rm d})$ & $5.66816\pm0.00005$ & $13.17552\pm0.00005$ & $21.77574_{-0.00004}^{+0.00006}$ & $31.7852\pm0.0002$ & $41.0274\pm0.0002$ \vspace{0.1cm} \\
$t_0\,({\rm BJD}-2454900)$ & $64.887\pm0.006$ & $76.679_{-0.003}^{+0.004}$ & $122.641_{-0.003}^{+0.004}$ & $68.859\pm0.005$ & $105.604\pm0.005$ \vspace{0.1cm} \\
$T_{\rm dur}\,({\rm hr})$ & $5.0\pm0.3$ & $6.7\pm0.2$ & $8.0\pm0.2$ & $8.7\pm0.2$ & $9.8\pm0.2$ \vspace{0.1cm} \\
$T_{\rm depth}\,({\rm ppt})$ & $0.086_{-0.004}^{+0.005}$ & $0.275\pm0.007$ & $0.801_{-0.009}^{+0.011}$ & $0.455\pm0.009$ & $0.573_{-0.008}^{+0.011}$ \vspace{0.1cm} \\
$\sqrt{e}\cos\omega$ & $0.0\pm0.2$ & $-0.03_{-0.07}^{+0.11}$ & $-0.05_{-0.04}^{+0.08}$ & $0.04_{-0.06}^{+0.05}$ & $0.01_{-0.05}^{+0.07}$ \vspace{0.1cm} \\
$\sqrt{e}\sin\omega$ & $0.0\pm0.2$ & $-0.02_{-0.08}^{+0.11}$ & $0.08_{-0.08}^{+0.04}$ & $-0.08_{-0.03}^{+0.07}$ & $0.08_{-0.09}^{+0.02}$ \vspace{0.1cm} \\
$e$ & $<0.2$ & $<0.05$ & $<0.03$ & $<0.02$ & $<0.02$ \vspace{0.1cm} \\
$b$ & $<0.5$ & $<0.4$ & $<0.4$ & $0.27_{-0.06}^{+0.07}$ & $0.17_{-0.11}^{+0.09}$ \vspace{0.1cm} \\
$i\,({\rm deg})$ & $>87.0$ & $>88.6$ & $>89.02$ & $89.4\pm0.1$ & $89.7_{-0.1}^{+0.2}$ \vspace{0.1cm} \\
$a/R_\star$ & $8.7_{-0.2}^{+0.1}$ & $15.3_{-0.3}^{+0.2}$ & $21.4_{-0.4}^{+0.3}$ & $27.6_{-0.5}^{+0.4}$ & $32.7_{-0.6}^{+0.4}$ \vspace{0.1cm} \\
$R_{\rm p}/R_\star\,(10^{-2})$ & $0.85\pm0.03$ & $1.51\pm0.02$ & $2.58\pm0.02$ & $1.96_{-0.02}^{+0.03}$ & $2.19\pm0.02$ \vspace{0.1cm} \\
$M_{\rm p}/M_\star\,(10^{-6})$ &  & $<46$ & $<20$ & $16.1_{-2.3}^{+2.7}$ & $19.8_{-2.6}^{+4.1}$ \vspace{0.1cm} \\
\hline
$a\,({\rm AU})$ & $0.0673_{-0.0012}^{+0.0004}$ & $0.1181_{-0.0020}^{+0.0008}$ & $0.165_{-0.003}^{+0.001}$ & $0.212_{-0.004}^{+0.001}$ & $0.252_{-0.004}^{+0.002}$ \vspace{0.1cm} \\
$F_{\rm p}\,(F_\oplus)$ & $697_{-28}^{+31}$ & $226.5_{-9.1}^{+10.2}$ & $115.9_{-4.7}^{+5.2}$ & $70.0_{-2.8}^{+3.1}$ & $49.8_{-2.0}^{+2.2}$ \vspace{0.1cm} \\
$R_{\rm p}\,(R_\oplus)$ & $1.54_{-0.05}^{+0.06}$ & $2.73\pm0.06$ & $4.67\pm0.09$ & $3.54_{-0.07}^{+0.09}$ & $3.96_{-0.07}^{+0.09}$ \vspace{0.1cm} \\
$M_{\rm p}\,(M_\oplus)$ &  & $<19$ & $<8.2$ & $6.6_{-1.0}^{+1.1}$ & $8.2_{-1.2}^{+1.6}$ \vspace{0.1cm} \\
$\rho_{\rm p}\,({\rm g/cm^3})$ &  & $<5.1$ & $<0.4$ & $0.8\pm0.1$ & $0.7\pm0.1$ \vspace{0.1cm} \\
\hline
$f_{\rm env}\,(\%)$ &  &  &  & $7.0_{-0.5}^{+0.6}$ & $10.3\pm0.6$ \vspace{0.1cm} \\
$M_{\rm c}\,(M_\oplus)$ &  &  &  & $6.2_{-0.9}^{+1.0}$ & $7.4_{-1.2}^{+1.3}$ \vspace{0.1cm} \\

		\noalign{\vskip0.5mm}
		\hline
	\end{tabular}
	\end{center}
\end{table*}

Considering the high-precision of $\rho_\star^{\rm tr}$, we also performed an MCMC fit using the CKS $T_{\rm eff}$, and ${\rm [Fe/H]}$ values along with $R_\star^{\rm EDR3}$ and the $\rho_\star^{\rm tr}$ constraints. This yielded the stellar parameter posterior distributions shown in Fig.~\ref{fig:HRD_corner}. We find that with the inclusion of $\rho_\star^{\rm tr}$, the resulting $R_\star$ posterior distribution inferred from the MIST models, which yields $R_\star=1.66\pm0.03\,R_\odot$, is notably narrower than that of either $R_\star^{\rm EDR3}$ or $R_\star^{\rm CKS}$ by a factor $\sim2$. In summary, we adopt the parameters derived using the CKS $T_{\rm eff}$, the CKS ${\rm [Fe/H]}$, $R_\star^{\rm EDR3}$, and $\rho_\star^{\rm tr}$ (reported in Table \ref{tbl:st_param}).

The position of Kepler-33 on the Hertzsprung-Russell Diagram (HRD) is shown in Fig.~\ref{fig:HRD_corner} (inset, top right). Based on our analysis, we find that Kepler-33 has a mass of $1.26_{-0.06}^{+0.03}\,M_\odot$ and an age of $4.2_{-0.3}^{+1.3}\,{\rm Gyr}$; it exhibits a fractional main sequence age of $0.93_{-0.05}^{+0.01}$ and is therefore evolving off of the main sequence as previously noted \citep[e.g.,][]{chachan2020}. Our results are consistent with the analysis carried out by \citet{lissauer2012}, who derive similar bimodal $M_\star$ and age posteriors (see their Fig. 5) characterized by two solutions near the terminal age main sequence: a low-$M_\star$/high-age solution ($\approx1.2\,M_\odot$ and $5.5\,{\rm Gyrs}$) along with a more-probable high-$M_\star$/low-age solution ($\approx1.3\,M_\odot$ and $4.2\,{\rm Gyrs}$). We found that adopting a flat age prior yields the same bimodality with a higher probability also being attributed to the younger solution.

\emph{Gaia} astrometry can also be used to obtain stellar age constraints by comparison of the star's kinematical properties with model predictions \citep{maciel2011}. \citet{almeida-fernandes2018} describe a method of deriving an age probability distribution function based on a star's peculiar velocities ($U$, $V$, and $W$). We calculated Kepler-33's $U$, $V$, and $W$ parameters using the \emph{Gaia} EDR3 position, proper motion, and parallax along with the barycentric systemic radial velocity of $\gamma_0=14.1\,{\rm km/s}$ reported by \citet{petigura2017}. Applying this $UVW$ method yields a kinematic age of $t_{\rm kin}=6.52_{-2.92}^{+5.35}\,{\rm Gyrs}$, which is consistent with either of the bimodal solutions shown in Fig. \ref{fig:HRD_corner} within $1\sigma$.

\section{Planet Properties}\label{sect:planets}

\subsection{Radii and masses}\label{sect:Mp_Rp}

\begin{figure}
	\centering
	\includegraphics[width=1\columnwidth]{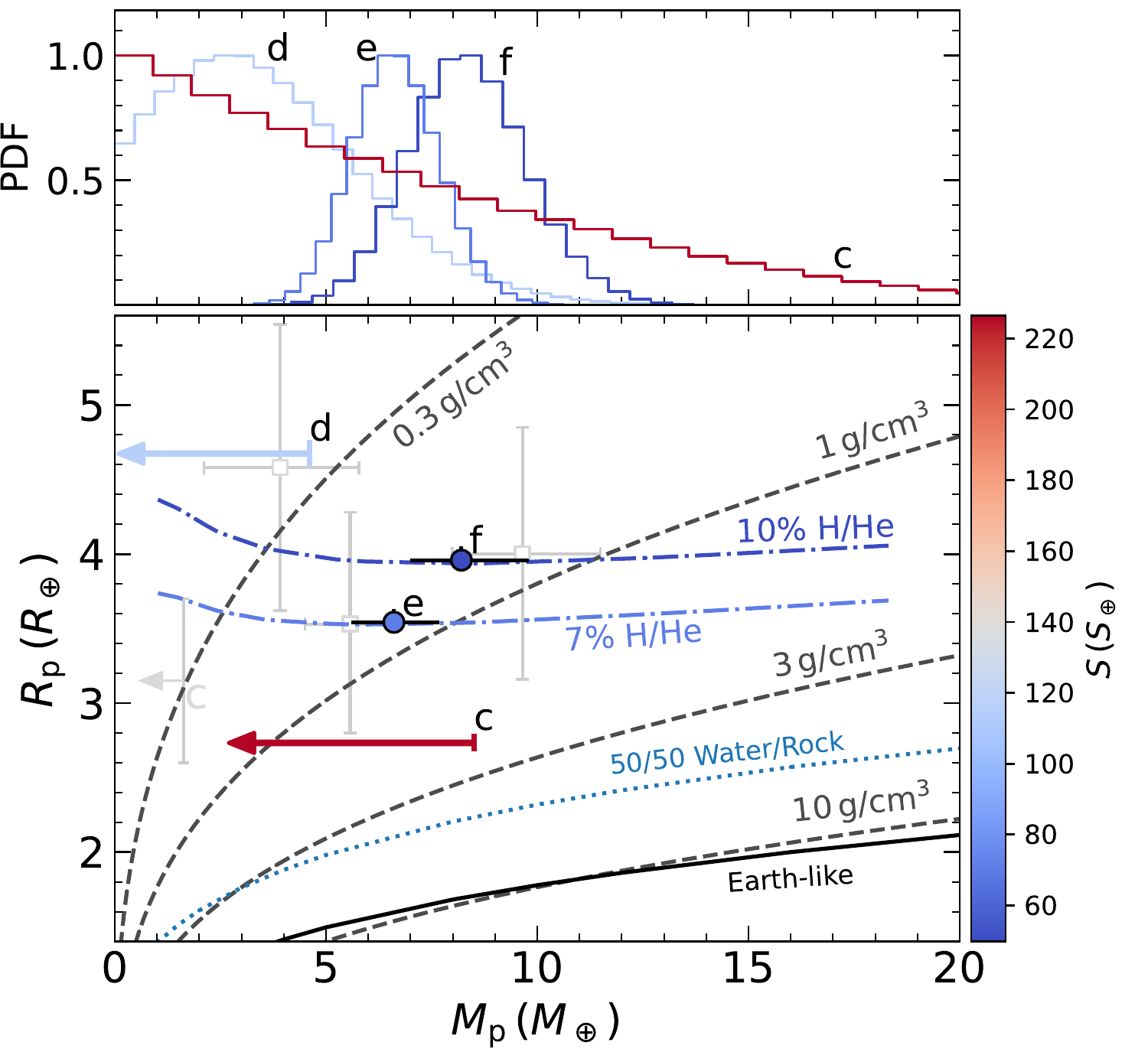}\vspace{-0.1cm}
	\caption{Colored circles show the masses and radii for Kepler-33 e and f derived using the adopted stellar parameters, where the color corresponds to each planet's insolation flux ($S$) (bottom panel). The $1\sigma$ upper mass limit of Kepler-33~c and d are denoted by the arrows. Gray squares/arrows indicate the masses and radii reported by \cite{hadden2016}. The colored dash-dotted lines are the best-fitting mass-radius relations for planets e and f derived using the models published by \cite{lopez2014}. The solid black and dotted blue lines show the two core mass-radius relations used in this work \citep{lopez2014,zeng2019}. The top panel shows the marginalized posterior distributions for the masses of planets c, d, e, and f.}
	\label{fig:Mp_Rp}
\end{figure}

The photodynamical modelling described in Sect.~\ref{sect:transit} yielded posterior distributions for, among other parameters, $R_{\rm p}/R_\star$ and $M_{\rm p}/M_\star$. Constraints for $R_{\rm p}$ and $M_{\rm p}$ were derived using the $R_\star$ and $M_\star$ posterior samples associated with the \texttt{isochrone} model fitting (Sect.~\ref{sect:star}) using (1) the adopted stellar posteriors derived using $T_{\rm eff}^{\rm CKS}$, ${\rm [Fe/H]}^{\rm CKS}$, $R_\star^{\rm EDR3}$, and $\rho_\star^{\rm tr}$ (Table \ref{tbl:st_param}) and (2) using the posteriors derived with only $T_{\rm eff}^{\rm CKS}$, ${\rm [Fe/H]}^{\rm CKS}$, and $R_\star^{\rm EDR3}$. The planet radii derived in the latter case are found to be larger by $\approx3\%$ (corresponding to $\approx0.4-0.6\sigma$ based on the lower/upper uncertainties estimated in both cases), while the masses of planets e and f (i.e., the two planets with well-constrained masses) are essentially identical. The masses and radii along with the mass posterior distributions derived for Kepler-33 c, d, e, and f using the adopted stellar parameters are shown in the mass-radius diagram in Fig.~\ref{fig:Mp_Rp}. As noted in Sect.~\ref{sect:transit}, no TTVs induced by planet b were detected, and only upper limits in $M_{\rm p}$ were obtained for planets c and d. The radii of planets d, e, and f are found to be consistent with values previously reported by \citet{hadden2016} within $0.2-1.1\sigma$ while c differs by $2.3\sigma$; the masses of planets e and f agree with the published values within $1.0\sigma$. The $R_{\rm p}$ uncertainties for all five planets are significantly reduced, which can largely be attributed to the precise parallax measurement provided by \emph{Gaia}; the estimated uncertainties in $M_{\rm p,e}$ and $M_{\rm p,f}$ are reduced by $\approx20-30\,\%$.

Densities of Kepler-33 e and f derived from the masses and radii are found to be consistent with being equal with one another within $0.2\sigma$ ($\rho_{\rm e}=0.8\pm0.1\,{\rm g/cm}^3$ and $\rho_{\rm f}=0.7\pm0.1\,{\rm g/cm}^3$). Planet d exhibits a notably lower density with an estimated $1\sigma$ upper limit of $0.25\,{\rm g/cm^3}$ -- significantly lower than the $0.37\,{\rm g/cm^3}$ upper bound reported by \citet{chachan2020} based on the mass measurements of \citet{hadden2016}. A precise constraint on the relative densities between each of the planets can be obtained using the posteriors derived directly from the transit modelling; we obtain $1\sigma$ upper limits for the ratio of densities of $\rho_{\rm d}/\rho_{\rm e}<0.34$ and $\rho_{\rm d}/\rho_{\rm f}<0.31$. As with planet d, only upper mass limits were derived for Kepler-33~c. 

\subsection{Composition}\label{sect:composition}

The precise radii and orbital periods derived for Kepler-33 d, e, and f place them well above the radius cliff \citep{hsu2019} strongly suggesting that they host atmospheres characterized by low mean molecular weights (i.e., $\mu\sim2.2$ for a predominantly H/He atmosphere). Planet c is positioned close to the cliff and well-above the radius valley \citep{fulton2017,vaneylen2018} suggesting that it exhibits a similar composition. This characterization is also consistent with their radii and bulk density constraints \citep[e.g.,][]{weiss2014a}: planets d, e, and f all exhibit $\rho<1\,{\rm g/cm^3}$ while planet c has $\rho<5.1\,{\rm g/cm^3}$, which requires light gases. No TTVs associated with Kepler-33~b were detected and thus, no mass constraints were obtained; however, the planet is positioned below the radius valley with $R_{\rm p}=1.54\,R_\oplus$ and $P_{\rm orb}=5.67\,{\rm d}$ suggesting that it is likely a rocky super-Earth.

Precise upper/lower radius constraints were obtained for all five of Kepler-33's planets while  the masses could be constrained relatively well only for planets e and f. For planets e and f, we compared the derived properties with publicly available model grids generated for low-mass planets hosting H/He envelopes. This allowed the envelope mass fraction ($f_{\rm env}\equiv M_{\rm env}/M_{\rm p}$ where $M_{\rm env}$ is the mass of the H/He envelope) associated with each planet to be derived. The core masses can also be calculated using the derived $f_{\rm env}$ values since $M_{\rm c}=(1-f_{\rm env})M_{\rm p}$.

Two planetary evolution model grids consisting of planets with solid cores (i.e., a metallic core and rocky mantle) surrounded by H/He envelopes were used for the analysis. The sub-Saturn planetary evolution models calculated by \citet{lopez2014} consist of $R_{\rm p}$ values given as a function of total mass ($1\leq M_{\rm p}/M_\oplus\leq20$), age ($0.1\leq{\rm Age/Gyr}\leq10$), insolation flux ($0.1\leq S/S_\oplus\leq1000$), and envelope mass fraction ($0.01\%\leq f_{\rm env}\leq20\%$). The models describe the thermal evolution of low-mass planets with H/He envelopes without including atmospheric mass-loss. Two grids are provided corresponding to a solar metallicity and an enhanced opacity ($50\times$ solar metallicity) where the latter is characterized by a slightly shorter cooling time. The two grids yielded essentially identical $f_{\rm env}$ values likely due to Kepler-33's advanced age; therefore, we report only those results derived using the solar metallicity grid.

Fitting of the interior composition with the pre-computed model grids was carried out using an MCMC analysis similar to that applied to the stellar evolutionary model fit (Sect.~\ref{sect:star}). We used the \texttt{emcee} Ensemble Sampler \citep{foreman-mackey2013a} with 50 walkers taking 25000 steps each, and then discarding the first 1000 steps as burn-in. For each step, the total planet radius is calculated from the adopted model grid by linearly interpolating across each grid's relevant parameter space and comparing with the measured radius. In the case of the \citet{lopez2014} grid, $t_{\rm age}$, $M_{\rm p}$, $F_{\rm p}$, and $f_{\rm env}$ are free parameters. Gaussian priors were adopted for $M_{\rm p}$, $F_{\rm p}$, and $t_{\rm age}$ defined using the previously derived values (Sections \ref{sect:star} and \ref{sect:Mp_Rp}) and the average of the lower/upper $1\sigma$ errors; a uniform prior was adopted for $f_{\rm env}$ defined by the grid limits. Using the \citet{lopez2014} grids to fit the measured stellar/planetary properties yield envelope mass fractions for Kepler-33~e and f of $7.0_{-0.5}^{+0.6}\%$ and $10.3\pm0.6\%$, respectively, and corresponding core masses of $6.2_{-0.9}^{+1.0}M_\oplus$ and $7.4_{-1.2}^{+1.3}M_\oplus$ (these $f_{\rm env}$ and $M_{\rm c}$ values are also reported in Table \ref{tbl:pl_param}). 

\section{Discussion}\label{sect:disc}

\begin{figure}
	\centering
	\includegraphics[width=1\columnwidth]{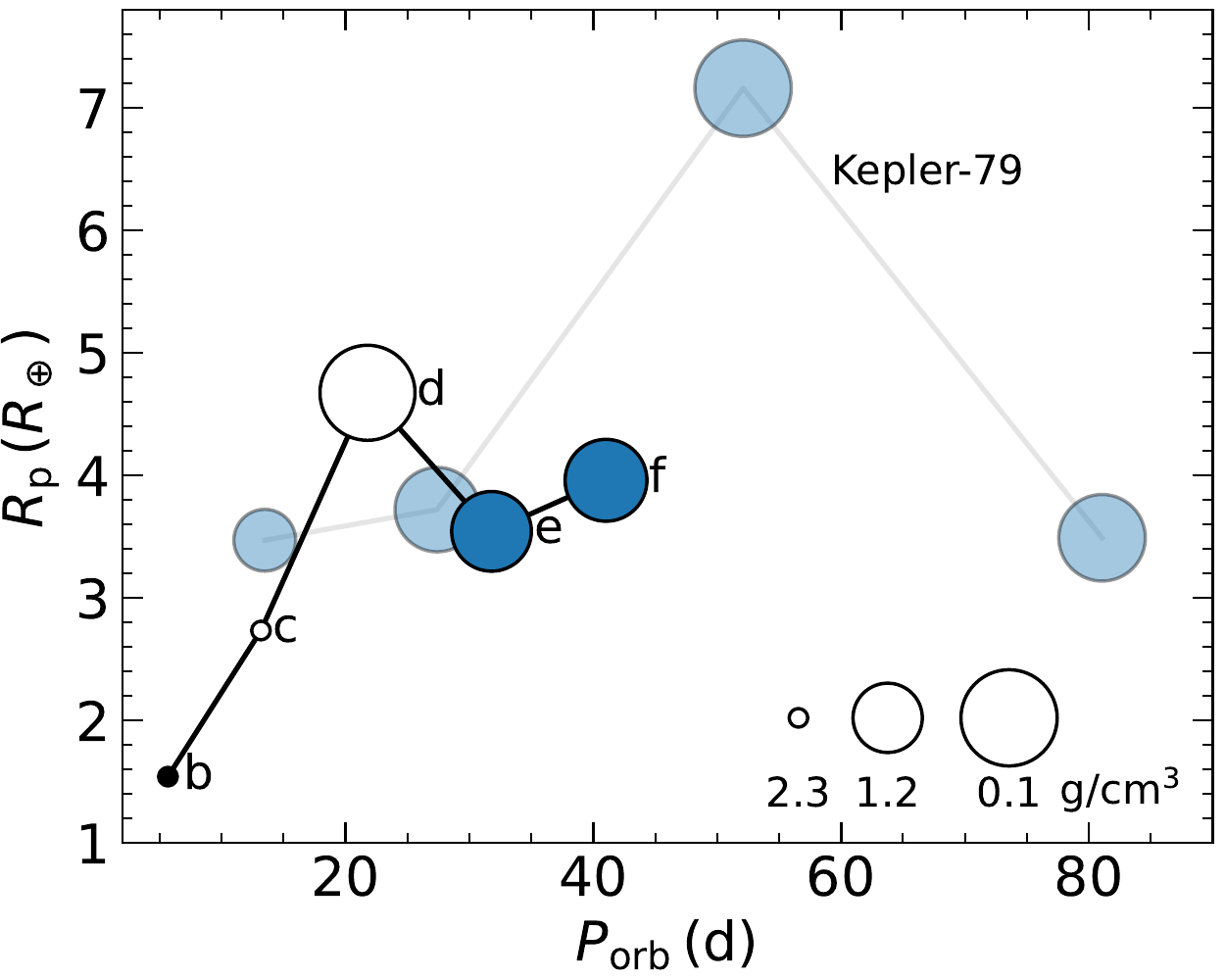}\vspace{-0.1cm}
	\caption{Orbital periods versus radii of Kepler-33's five planets. The size of each circle increases with decreasing density (i.e., larger circles are puffier planets); open circles correspond to upper $\rho$ limits, blue circles have well-constrained $\rho$, while no $\rho$ constraints were derived for the black circle (planet b). Kepler-79 is shown for comparison \citep{jontof-hutter2014}.}
	\label{fig:arch}
\end{figure}

Based on their core accretion models, \citet{lee2016} predict that, within a given system, planets found further from their host-stars are likely to have lower bulk densities and higher envelope mass fractions relative to their shorter period companions. Similar to Kepler-79, in which planet d interior to e has a lower density \citep{jontof-hutter2014}, Kepler-33 is found to be inconsistent with this predicted sequence, since planet d ($a=0.165\,{\rm AU}$, and a $2\sigma$ upper limit of $\rho<0.4\,{\rm g/cm^3}$) is notably puffier (lower in density) than either planets e ($a=0.212\,{\rm AU}$, $\rho=0.8\pm0.1\,{\rm g/cm^3}$) or f ($a=0.252\,{\rm AU}$, $\rho=0.7\pm0.1\,{\rm g/cm^3}$) (Fig. \ref{fig:arch}). This potential conflict is also apparent when comparing the present-day envelope mass fractions of $f_{\rm env,e}=7.0_{-0.5}^{+0.6}\%$ and $f_{\rm env,f}=10.3\pm0.6\%$ with those estimated for planet d: only upper limits on planet d's mass were able to be derived ($M_{\rm p,d}<8.2\,M_\oplus$ corresponding to $2\sigma$), however, assuming that $M_{\rm p,d}=3\,M_\oplus$ or $M_{\rm p,d}=5\,M_\oplus$ yields $f_{\rm env,d}$ values of $\approx12\%$ and $\approx13\%$, respectively.

It is plausible that outer planets with lower radii than their shorter period companions may in some cases undergo more substantial mass-loss due to giant impacts \citep[e.g.,][]{schlichting2018,kegerreis2020a}. Applying this scenario to Kepler-33 (as well as Kepler-79) could potentially explain the discrepancy between the predicted and the observed density trends. Another potential explanation for Kepler-33~d's large radii is that magma freezing within the cooling core may have released volatiles into the atmosphere \citep{kite2020a} causing the planet to be re-inflated \citep{elkins-tanton2011}. The potential magnitude of this effect on the observed radius of planet d is uncertain; however, work by \citet{schlichting2022a} suggests that only a negligible fraction of the planet's total H$_2$ can be stored within a magma ocean for sub-Neptunes with H/He envelope mass fractions of $f_{\rm env}\gtrsim2\%$, implying that the impact on the radius of planets d, e, and f may be insignificant.

Other mechanisms have also been proposed to explain the existence of puffy planets like Kepler-33~d. The presence of high-altitude photochemical hazes have been shown to plausibly enhance the observed radii of low-mass planets \citep[e.g.,][]{lammer2016,gao2020a} thereby yielding lower-than-expected bulk densities. Including additional heat sources in planetary evolution models has also been shown to lead to larger planet radii \citep[e.g.,][]{vazan2018}. For instance, \citet{millholland2019} show that obliquity tides may heat the interiors of planets that have near-resonance orbits causing the envelope radius to be significantly increased.

\subsection{Future observations}

\subsubsection{Mass refinement}
While we derived relatively precise constraints on the masses of planets e and f, our analysis was only able to yield upper limits for planets c and d and no useful mass constraints for planet b. Based on the derived posteriors, we find that planet d is expected to exhibit a radial velocity semi-amplitude of $<0.8\,{\rm m/s}$; achieving the necessary sensitivity to detect such a signal is currently challenging but may be feasible with extreme-precision radial velocity measurements obtained using instruments such as the Keck Planet Finder \citep{kassis2018}.

\begin{figure}
	\centering
	\includegraphics[width=1\columnwidth]{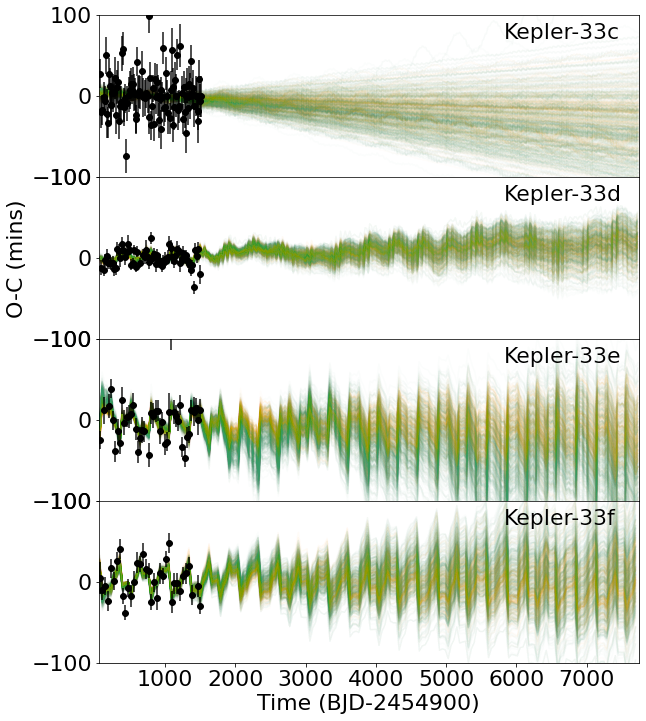}\vspace{-0.0cm}
	\includegraphics[width=0.95\columnwidth]{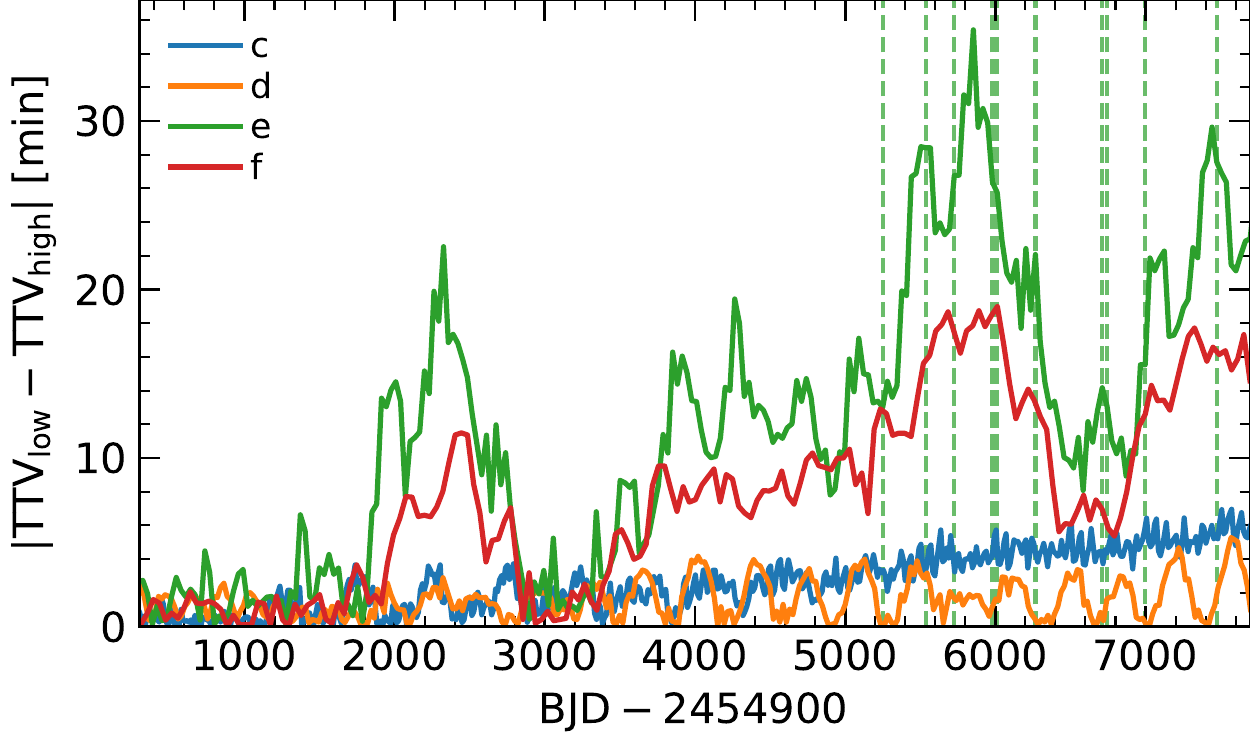}
	\caption{\textbf{Top:} O$-$C diagram showing the predicted TTVs from ${\rm BJD}\approx2456400$ (April 2013) to ${\rm BJD}\approx2462600$ (April 2030). Colors correspond to low-mass ($<2\,M_\oplus$, yellow) and high-mass ($>2\,M_\oplus$, green) solutions for planet d. \textbf{Bottom:} Estimated differences between the predicted TTVs for planets c, d, e, and f if the mass of planet d is $<2\,M_\oplus$ (TTV$_{\rm low}$) or $>2\,M_\oplus$ (TTV$_{\rm high}$). Vertical dashed lines indicate transits of planet e that are observable from Palomar Observatory. }
	\label{fig:ttv_predicted}
\end{figure}

Alternatively, a more feasible approach to refining the masses obtained here -- particularly that of planet d -- may involve observing additional transits (i.e., additional TTVs). In order to evaluate the utility of such observations, we carried out $N$-body simulations using the posteriors derived from the photodynamics analysis, which allow future transit times to be predicted. In Fig.~\ref{fig:ttv_predicted} (top), we show the predicted TTVs for planets c, d, e, and f up to Feb. 2031 (${\rm BJD}\approx2462900$) assuming low- and high-mass solutions for planet d ($<2\,M_\oplus$ and $>2\,M_\oplus$, respectively). Fig.~\ref{fig:ttv_predicted} (bottom) shows the estimated differences between the predicted TTVs in the low- and high-mass median solutions associated with each planet. We find that these two scenarios are most easily distinguished by obtaining future transit observations of planet e, which exhibit the largest differences ($\lesssim30\,{\rm min}$).

Detecting TTVs associated with Kepler-33's five transiting planets is not feasible using observations obtained by the TESS mission \citep{ricker2014} due to the star's low brightness ($K{\rm p}=14.1\,{\rm mag}$, $J=12.9\,{\rm mag}$) \citep[e.g.,][]{hadden2019}. However, the upcoming {\it PLATO} mission \citep{rauer2014}, currently scheduled for launch in 2026, may be suitable. Transits of planets e and f are expected to be detectable using ground-based instruments such as the Wide-field InfraRed Camera installed on the 5.1~\,{\rm m} Hale Telescope at Palomar Observatory \citep{wilson2003} \citep[see Fig. 3 of][]{vissapragada2020b}. Transits for planet e that are observable from Palomar having mid-point times coincident with low airmasses of $<1.3$ are shown in Fig. \ref{fig:ttv_predicted} (top).

\subsubsection{JWST transmission spectra}
Based on their derived masses, radii, and orbital periods, Kepler-33~d, e, and f likely host thick atmospheres that may be suitable for detailed characterization using \emph{JWST}. Of the three planets, planet d exhibits the largest transmission spectroscopy metric (${\rm TSM}\gtrsim29$), which provides an indication of the expected signal strength \citep{kempton2018}.

\begin{figure}
	\centering
	\includegraphics[width=1\columnwidth]{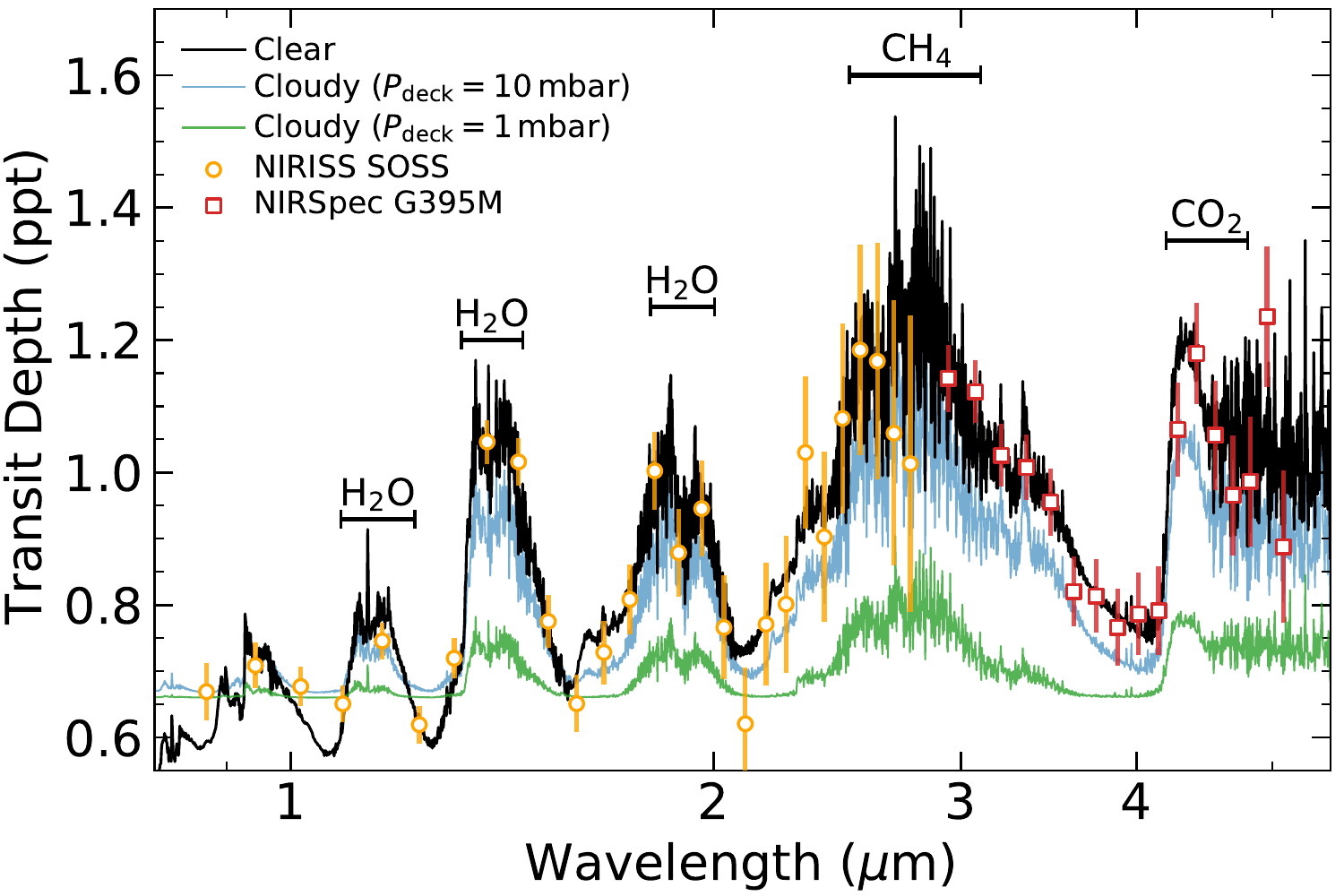}\vspace{-0.0cm}
	\caption{Simulated \emph{JWST} observations of Kepler-33~d assuming $M_{\rm p,d}=4\,M_\oplus$ and $T_{\rm eq}=908\,{\rm K}$. Model transmission spectra for one clear (black) and two cloudy atmospheres (blue and green) are shown. The simulated NIRISS and NIRSpec observations (yellow circles and red squares, respectively) are calculated for three transits using the clear atmosphere.}
	\label{fig:jwst}
\end{figure}

In Fig. \ref{fig:jwst}, we show simulated \emph{JWST} transmission spectra for Kepler-33~d. The model atmospheres used to generate the measurements were calculated using petitRADTRANS \citep{molliere2019} assuming a solar metallicity (${\rm [Fe/H]}=0$), a solar C/O ratio (${\rm C/O}=0.55$), and a planet mass of $4\,M_\oplus$. We adopt an isothermal atmosphere with a temperature equal to Kepler-33~d's equilibrium temperature of $908\,{\rm K}$ (calculated with an albedo of zero and assuming full heat redistribution). The models include molecular absorption from H$_2$O, CO, CH$_4$, CO$_2$, and NH$_3$ whose abundances were estimated assuming chemical equilibrium \citep{molliere2017}; Rayleigh scattering due to H$_2$ and He and collision induced absorption due to H$_2$-H$_2$ and H$_2$-He interactions are also included in the model. Three model spectra are shown in Fig. \ref{fig:jwst}: one for a clear atmosphere free of clouds/hazes (black line) and two with gray cloud decks at pressures of $10\,{\rm mbar}$ (blue line) and $1\,{\rm mbar}$ (green line). The high-altitude cloud deck at $1\,{\rm mbar}$ was chosen based on modelling of observed transmission spectra of Kepler-51~b and d \citep{libby-roberts2020}. The simulated NIRISS SOSS and NIRSpec G395M \emph{JWST} observations shown in Fig. \ref{fig:jwst} (yellow circles and red squares, respectively) were calculated using PandExo \citep{batalha2017} assuming three transits.

The simulated measurements shown in Fig. \ref{fig:jwst} suggest that, in the absence of high-altitude clouds, key molecular absorption features such as that of water \citep[e.g.][]{benneke2019a} can potentially be detected from \emph{JWST} observations of Kepler-33~d. Such observations would be particularly useful if the mass constraints derived here are improved with RV/TTV measurements.

\section{Summary}\label{sect:summary}

The updated constraints of Kepler-33's stellar and planetary properties derived in this work provide a clearer picture of this highly compact system. Combining the \emph{Gaia} EDR3 catalog with the temperature and metallicity constraints reported for the California-\emph{Kepler} Survey \citep{fulton2018}, we find that the host-star is $4.2_{-0.3}^{+1.3}\,{\rm Gyrs}$ old and will soon be evolving off the main sequence. The photodynamics analysis presented here confirms previous findings that Kepler-33 e and f ($R_{\rm p}\sim3.5-4.0\,R_\oplus$) exhibit bulk densities $\sim0.7\,{\rm g/cm^3}$ while, contrary to \citet{hadden2016}, only an upper mass of d is obtained ($M_{\rm p,d}<4.6-8.2\,M_\oplus$ corresponding to $1\sigma$ and $2\sigma$, respectively). The mass of planet d implies a relatively low density of $\lesssim0.4\,{\rm g/cm^3}$. Further refinement of Kepler-33~d's mass is necessary to determine whether the planet's density is comparable to super-puffs like Kepler-79~d ($0.08\pm0.02\,{\rm g/cm^3}$) and Kepler-51~b ($0.06\pm0.03\,{\rm g/cm^3}$) \citep{masuda2014} and whether it's age and density can be reconciled with theoretical predictions of sub-Neptune/sub-Saturn mass-loss and core-accretion predictions.

\begin{acknowledgments}
We would like to thank the anonymous referee for providing valuable and detailed feedback that significantly improved this work. Furthermore, we thank William Borucki and Jacob Kegerreis for taking the time to review the manuscript and for providing very helpful comments and suggestions.

J.F.R. acknowledges research funding support from the Canada Research Chairs program and NSERC Discovery Program. This research was enabled, in part, by support provided by Calcul Québec (\url{www.calculquebec.ca}) and ComputeCanada (\url{www.computecanada.ca})

\end{acknowledgments}

\software{isoclassify \citep{huber2017,berger2020}, MIST \citep{choi2016}, mwdust \citep{bovy2016}, TRANSITFIT \citep{rowe2015,rowe2016}, Mercury6 \citep{chambers1999}, isochrones \citep{morton2016}.}

The {\it Kepler} data used in this paper can be found in MAST: \dataset[10.17909/T9059R]{http://dx.doi.org/10.17909/T9059R}.

\bibliography{Kepler33}{}
\bibliographystyle{aasjournal}

\clearpage

\section*{Appendix}

\begin{figure*}[!h]
	\centering
	\includegraphics[width=1\textwidth]{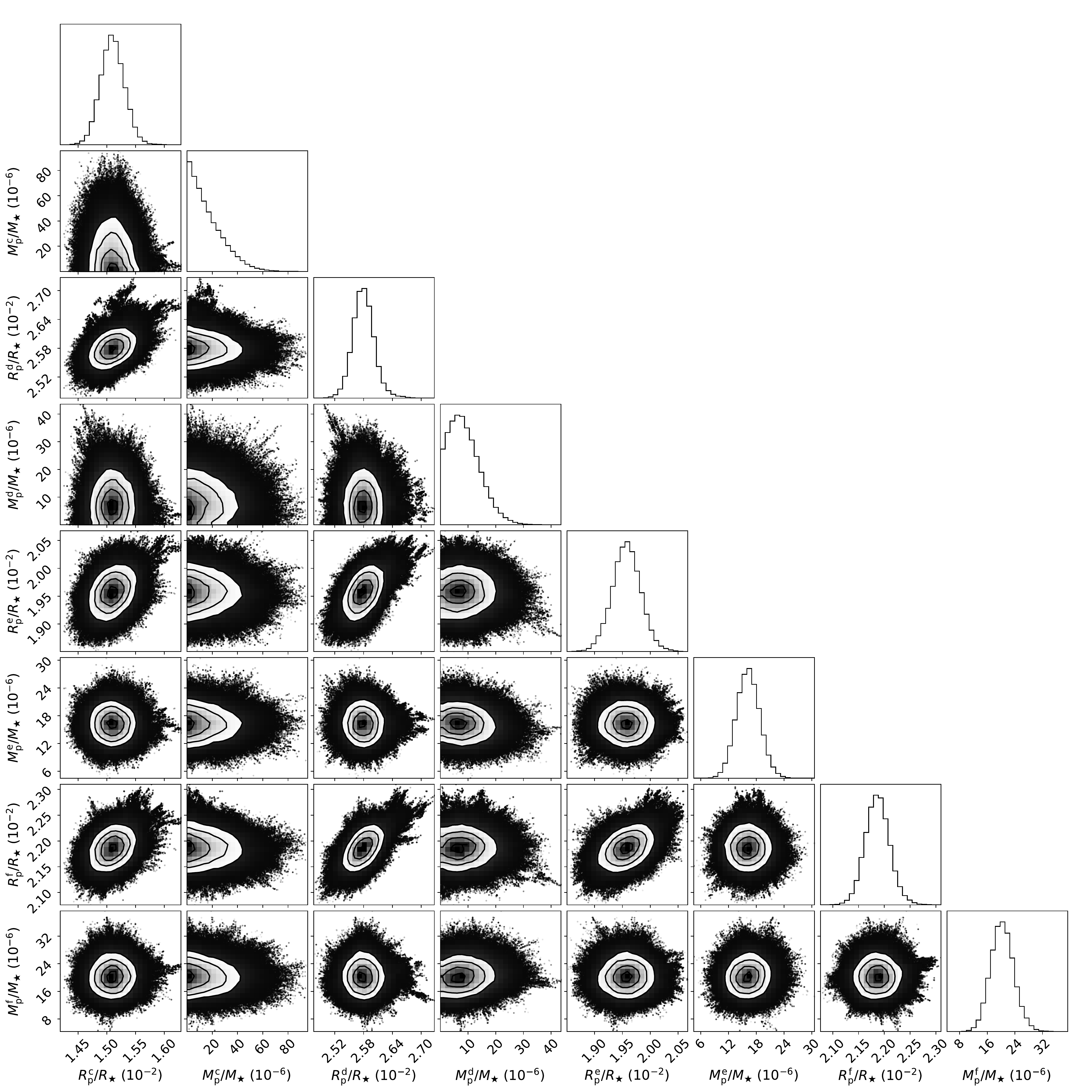}\vspace{-0.1cm}
	\caption{Marginalized posterior distributions of $R_{\rm p}/R_\star$ and $M_{\rm p}/M_\star$ for planets c, d, e, and f derived from the photodynamics modelling (Sect.~\ref{sect:transit}).}
	\label{fig:RpRs_MpMs_corner}
\end{figure*}

\begin{figure*}
	\centering
	\includegraphics[width=1\textwidth]{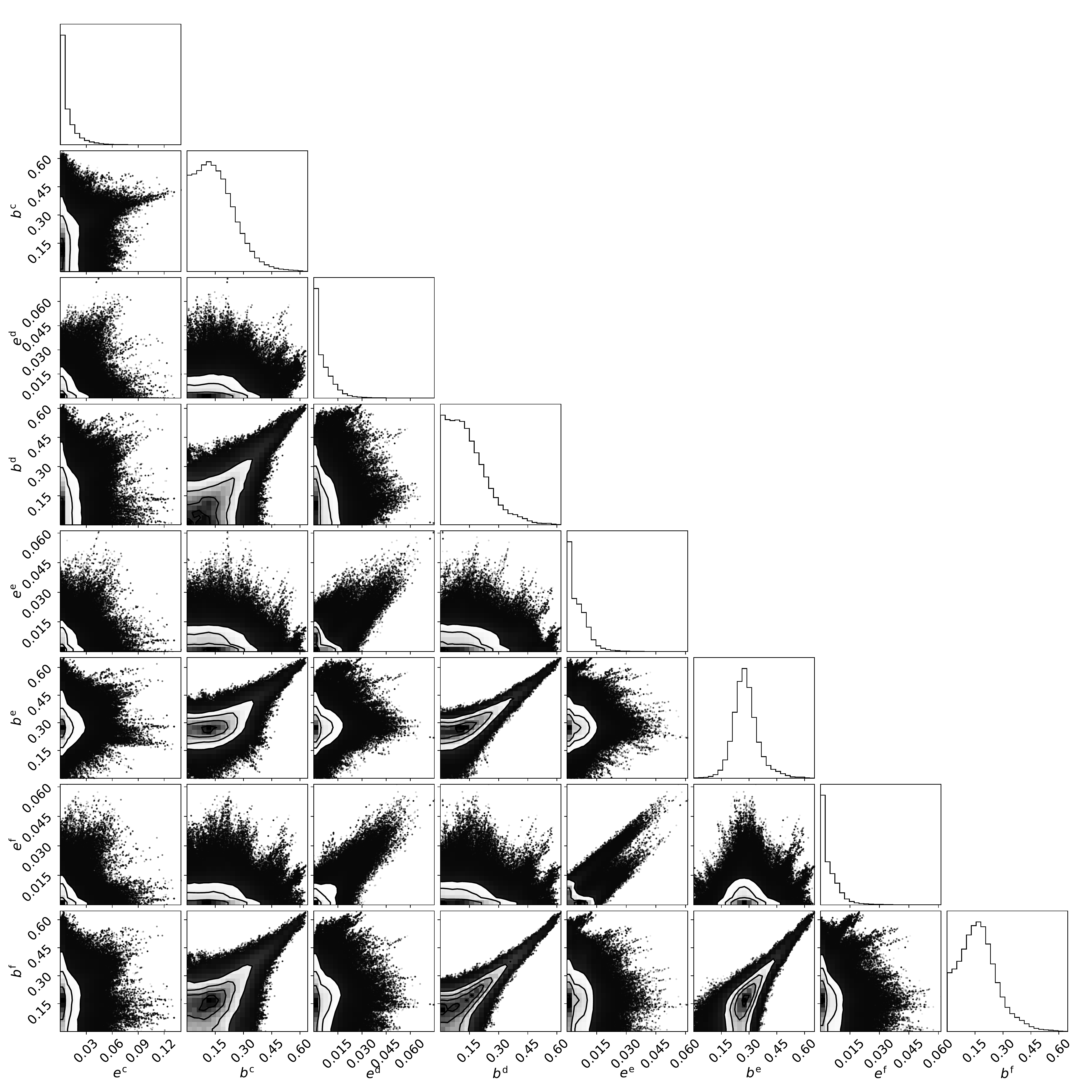}\vspace{-0.1cm}
	\caption{Same as Fig.~\ref{fig:RpRs_MpMs_corner} but for impact parameters ($b$) and eccentricities ($e$).}
	\label{fig:b_corner}
\end{figure*}

Corner plots showing the marginalized posterior distributions associated with the derived radii, masses, impact parameters, and eccentricities are shown in Figures \ref{fig:RpRs_MpMs_corner} and \ref{fig:b_corner}.

\clearpage

\end{document}